\documentclass[%
 reprint,
superscriptaddress,
 amsmath,amssymb,
 aps,
]{revtex4-1}
\usepackage{float}
\usepackage{graphicx}% Include figure files
\usepackage{dcolumn}% Align table columns on decimal point
\usepackage{bm}% bold math
\usepackage{hyperref}% add hypertext capabilities
\usepackage{caption} % 在导言区加载 caption 宏包

\begin{document}

\preprint{APS/123-QED}

\title{
Testing Extended Theories of Gravity via Black Hole Photon Rings
}%

%\collaboration{MUSO Collaboration}%\noaffiliation

%\author{Charlie Author}
 %\homepage{http://www.Second.institution.edu/~Charlie.Author}

%\author{Delta Author}

%\collaboration{CLEO Collaboration}%\noaffiliation
\author{Qiao Yue}
% \altaffiliation{College of Physics, Guizhou University, Guiyang 550025, China}%Lines break automatically or can be forced with \\
\author{Zhaoyi Xu}

 \author{Meirong Tang}
 \email{Electronic address: tangmr@gzu.edu.cn(Corresponding author)
}
% \email{Second.Author@institution.edu}
\affiliation{%
College of Physics, Guizhou University,\\
 Guiyang 550025, China
}%
\date{\today}% It is always \today, today,
             %  but any date may be explicitly specified

\begin{abstract}

As a unique observational feature in strong gravitational fields, photon rings around black holes provide a method to evaluate general relativity and alternative gravity theories. This research delves into the optical characteristics of stationary, spherically symmetric black holes. These black holes follow the Konoplya-Zhidenko deformation rule in arbitrary gravity theories. This black hole is characterized by three independent parameters ($\varepsilon$, \(a_2\), \(b_2\)) to describe its divergence from the Schwarzschild geometry. This paper derives the geodesic equation and effective potential function from the Lagrangian, determines the photon sphere radius (unstable orbit), and analyzes the null geodesics using the reverse ray-tracing technique. This research finds that the effects of \(a_2\) and \(b_2\) on photon orbital dynamics exhibit observational degeneracy, while \(\varepsilon\) significantly governs photon capture characteristics. As \(\varepsilon\) increases, the radius of the photon sphere \(r_{\text{ph}}\) and the critical impact parameter \(b_{\text{ph}}\), and the innermost stable circular orbit radius \(r_{\text{isco}}\) all increase. The event horizon \( r_{\text{h}} \) corresponds to that of the Schwarzschild black hole, while the impact parameter range for the lens and photon rings is reduced. Black hole shadow and photon ring analyses across three emission models show that increasing \(\varepsilon\) shifts the peak rightward while enlarging the photon ring radius. The closer \(\varepsilon\) is to zero, the more the results approach the Schwarzschild case, the more the results approach the Schwarzschild case. Additionally, by combining EHT observational data on the shadow diameters of M87 and Sgr A*, we imposed constraints on the correlation parameter \(\varepsilon\) in the theoretical model(at the confidence level anchored by \(d_{\text{sh}}^{(M87^*)}\), the parameter \(\varepsilon\) is confined to the interval \(-0.09 \lesssim \varepsilon \lesssim 0.19\). For \(d_{\text{sh}}^{(Sgr A^*)}\), the constraint on \(\varepsilon\) is delineated as \(-0.280 \lesssim \varepsilon \lesssim 0.047\)). The results show that within the observationally allowed range of \(\varepsilon\) (such as \([-0.04, 0.04]\)), the characteristics of the black hole exhibit specific regularities with changes in \(\varepsilon\). In summary, the photon rings of this type of black hole do not exhibit degeneracy, and can theoretically distinguish different spacetime metrics, providing a new approach to constrain extended theories of gravity. 

\begin{description}
\item[Keywords]
Photon ring, Geodesic, Extended theories of gravity, Event Horizon Telescope (EHT)
\end{description}
\end{abstract}
%\keywords{Suggested keywords}%Use showkeys class option if keyword

\maketitle

%\tableofcontents
\section{Introduction}\label{1.0}

As extreme gravitational objects predicted by general relativity, black holes provide a natura ``strong-field laboratory'' for testing classical gravitational theories and exploring quantum gravitational effects through the spacetime geometry in their strong-field regions. While general relativity's weak-field predictions (planetary precession, gravitational redshift) have been experimentally confirmed with high precision \cite[e.g.][]{Will:2014kxa,Shapiro:1971iv, MisnerThorne2017Gravitation,Wald1984GeneralRelativity}, however, near event horizons (EHs) where curvature reaches \(10^{44} \, \text{cm}^{-2}\), spacetime structure might differ substantially from general relativity's predictions \cite[e.g.][]{Berti:2015itd,Tian:2019yhn,Horava:2009uw,cuk2016dynamical}. The Event Horizon Telescope (EHT) acquired the initial shadow images of M87*'s supermassive black hole in 2019 \cite{EventHorizonTelescope:2019dse,EventHorizonTelescope:2019uob,EventHorizonTelescope:2019jan,EventHorizonTelescope:2019ths,EventHorizonTelescope:2019pgp,EventHorizonTelescope:2019ggy}. These observations enable direct examination of supermassive black hole EHs with electromagnetic waves, converting this theoretical boundary into a physical phenomenon testable through repeated astronomical measurements. In 2022, detailed images of Sagittarius A*, our galaxy's central supermassive black hole, were published \cite{EventHorizonTelescope:2022wkp}. These images provide a natural laboratory for studying black hole physics and testing relativity, marking humanity's entry into an era of direct strong-field spacetime exploration through photon rings and shadows.

Photon rings are an intuitive manifestation of the dynamics of null geodesics in strong gravitational fields, essentially representing a multiply nested radiative structure formed by photons orbiting in unstable closed trajectories around black hole \cite[e.g.][]{Gralla:2019xty,Wang:2023vcv,Wang:2022yvi,Chael:2021rjo}. This structure, associated with black hole mass and spin, exists near the event horizon (EH) \cite[e.g.][]{Gogoi:2024vcx,Johnson:2019ljv,Himwich:2020msm}. Through observational analysis of photon rings, deeper insights into black hole accretion disk structure and gravitational lensing have been gained, Mass distribution in the vicinity of black holes, and diverse extended theories of gravitation \cite[e.g.][]{Beckwith:2004ae,Gao:2023mjb,Luminet:1979nyg,Gan:2021xdl,Takahashi:2004xh,Yang:2024utv,Zeng:2025kqw,Wang:2025ihg,Zeng:2024ptv,Zeng:2023zlf,Zeng:2022pvb,Zeng:2022fdm}. This confirms general relativity in strong fields while creating opportunities to study near-horizon spacetime, enhancing our grasp of gravity and curvature under extreme conditions. Since EHT's 2019 first black hole image release, detailed photon ring studies have emerged as a breakthrough area in astronomical research \cite[e.g.][]{Wu:2025hcu,Johannsen:2013vgc,He:2025rjq,Zeng:2021dlj}. This continuous and in-depth research has greatly improved the observational accuracy and depth of understanding of black holes in the astronomical community. Additionally, research on photon rings helps address core inquiries in physics, for instance, advancing the development of quantum gravity theories by providing critical experimental and observational evidence for their construction, and promoting the refinement of physical theories under extreme conditions \cite[e.g.][]{Zare:2024dtf,Mizuno:2018lxz,Heydari-Fard:2023ent}. At the same time, as a key feature of black holes, photon rings can help scientists understand the interactions between black holes and surrounding matter, as well as the impact of such interactions on the structure and evolution of the universe.

Extended gravity theories address general relativity's shortcomings at both infrared and ultraviolet scales \cite{Capozziello:2011et}. These theories enhance Einstein's framework by incorporating higher-order curvature terms ($R^{2}$, $R_{\mu \nu}R^{\mu \nu}$) and variably coupled scalar fields into the gravitational Lagrangian. These corrections derive from quantum gravity's effective action \cite{Baghram:2007df,Anderson:1971dm}. In cosmology, dark energy and dark matter phenomena \cite{Bassett:2002fe}, cannot be satisfactorily explained by general relativity, prompting scientists to seek more complete theories. As one of the early important extended theories of gravity \cite{Barros:1997qt}, the Brans-Dicke theory's action includes terms related to the metric and scalar fields. By introducing scalar fields to realize Mach's principle, this theory has profoundly influenced the development direction of subsequent extended gravity theories \cite{Brans:1961sx}. Since then, various extended theories of gravity have emerged. Among them, the f(R) gravity theory is a significant representative. In the metric formulation, the field equations of this theory are fourth-order, exhibiting greater complexity than those in general relativity, and can provide richer solutions, making it of great importance in cosmology \cite{DeFelice:2010aj,Sotiriou:2008rp,Kainulainen:2007bt}. Scalar-tensor gravity theories also fall into this category. These theories link scalar fields with geometry through non-minimal coupling and are widely used in theoretical research and the construction of cosmological models \cite{Damour:1992we,Sotiriou:2006hs}. In addition, there are also hybrid high-order-scalar-tensor gravity theories that combine non-minimal coupling and high-order terms \cite{Futamase:1989hb}. Furthermore, extended theories of gravity exhibit natural inflationary behavior in cosmology, successfully addressing limitations in general relativity's standard cosmological model. Their related models are highly consistent with observations of cosmic microwave background radiation, providing more powerful theoretical tools for studying the evolution and structure formation of the universe \cite{Capozziello:2007ec,Barrow:1988xh}. Meanwhile, these theory also provides an intermediate working scheme for the study of quantum gravity. Although it is not a complete quantum gravity theory \cite{Vilkovisky:1992pb}, it holds significant reference value in the process of developing toward it.

Next, this paper uses a general static spherically symmetric black hole model applicable across various gravitational theories, which follows the Konoplya-Zhidenko deformation rule. By performing precise calculations of geodesics in strong gravitational fields via reverse ray-tracing techniques and analyzing the optical appearance of this black hole in three toy emission models, our work aims to constrain extended theories of gravity. These studies reveal the unique advantages of photon rings as strong-field gravitational probes and provide new tools for theoretically interpreting future high-resolution black hole imaging data. In-depth research on photon rings and general black hole models following the Konoplya-Zhidenko deformation rule not only deepens our understanding of predictions from extended gravity theories but also lays a foundation for constraining new physical mechanisms — such as dark matter coupling and quantum gravitational effects — through observations.

This paper is organized as follows: Section \ref{2.0} introduces a general black hole model with static spherical symmetry in arbitrary gravitational theories, which follows the Konoplya-Zhidenko deformation rule. We analyze how deformation parameters $\varepsilon$ and $a_2$ influence the metric functions and EH. In Section \ref{3.0}, we first derive null geodesics and effective potential equations for this black hole. We then discuss the impact of parameters \(\varepsilon\) and \(a_2\) on the effective potential function is analyzed, and the EH radius, unstable photon sphere radius, and critical impact parameter are obtained. In Section \ref{4.0}, first, using null geodesic equations with an optically thin accretion disk model, we show how bending angles vary with impact parameters under different deformation parameters $\varepsilon$ and $b_2$, along with corresponding trajectories. We then constrain \(\varepsilon\) using EHT shadow diameter observations of M87* and Sgr A*. Finally, we present three transfer function relationships and compare emission intensity, observed intensity, and optical features across three emission models. Section \ref{5.0} concludes the paper. Additionally, throughout this work, geometric units are consistently employed, with $C=G=M=1$ and metric signature $(-, +, +, +)$.

\section{General black hole with Konoplya-Zhidenko deformation}\label{2.0}

In this part, we initially offer a concise overview of the generalized black hole derived from the modification of the Schwarzschild spacetime. This black hole results from general parameterization of spherically symmetric, asymptotically flat black holes across gravitational theories. In spherical coordinates, its metric tensor is \cite{Konoplya:2023owh}
\begin{equation}\label{1}
ds^2=-N^2(r)dt^2+\frac{B^2(r)}{N^2(r)}dr^2+r^2(d\theta ^2+sin^2\theta d\phi ^2), 
\end{equation}
where
\begin{equation}\label{2}
N^2(r)=(1-\frac{r_0}{r})(1-\frac{r_0\varepsilon }{r}-\frac{r_0^2\varepsilon }{r^2}+\frac{r_0^3}{r^2 }\frac{a_1}{r+a_2(r-r_0)}),
\end{equation}
\begin{equation}\label{3}
B^2(r)=(1+\frac{r_0^2}{r}\frac{b_1}{r+b_2(r-r_0)})^2.
\end{equation}
In this expression, \(N\) and \(B\) represent functions of the radial coordinate r, with \(r = r_0 > 0\) denoting the position of the EH. This metric depicts spherically symmetric spacetime using five parameters \(\{\varepsilon, a_1, a_2, b_1, b_2\}\) that quantify Schwarzschild deviations. It is important to emphasize that these five deformation parameters are not mutually independent. For proper weak-field Schwarzschild approximation, these deformation parameters must meet specific conditions
\begin{equation}\label{4}
a_1=-(3+a_2)\varepsilon, 
\end{equation}
\begin{equation}\label{5}
b_1=-\frac{4(2+a_2)(3+b_2)}{(3+a_2)^2}\varepsilon.    
\end{equation}
Therefore, the metric has only three independent deformation parameters \(\{\varepsilon, a_2, b_2\}\). Parameter \(\varepsilon\) quantifies the deviation of \(r_0\) from the Schwarzschild radius \(r = 2M\), while \(a_2\) and \(b_2\) define near-horizon geometry. To better simulate the Schwarzschild spacetime, this paper sets \(r_0 = 2M\). When deformation parameters are zero, the metric reduces to the Schwarzschild metric.

To preserve positive definiteness for metric coefficients \(N^2(r)\) and \(\frac{B^2(r)}{N^2(r)}\) beyond the EH and eliminate potential naked singularities, these parameters must comply with the following restrictions \cite{Xu:2024cfe}
\begin{equation}\label{6}
\begin{split}
a_2> -1,b_2>-1,\\  
a_1>2\varepsilon -1,1+b_1>0.
\end{split}  
\end{equation}
Therefore, the possible EH under this metric should only be a function of r and must satisfy
\begin{equation}\label{7}
g^{rr}=0, 
\end{equation}
that is
\begin{equation}\label{8}
N^2(r)=(1-\frac{r_0}{r})(1-\frac{r_0\varepsilon }{r}-\frac{r_0^2\varepsilon }{r^2}-\frac{r_0^3}{r^2}\frac{(3+a_2)}{r+a_2(r-r_0)}\varepsilon ) )=0.
\end{equation}

\begin{figure*}[]
\includegraphics[width=1\textwidth]{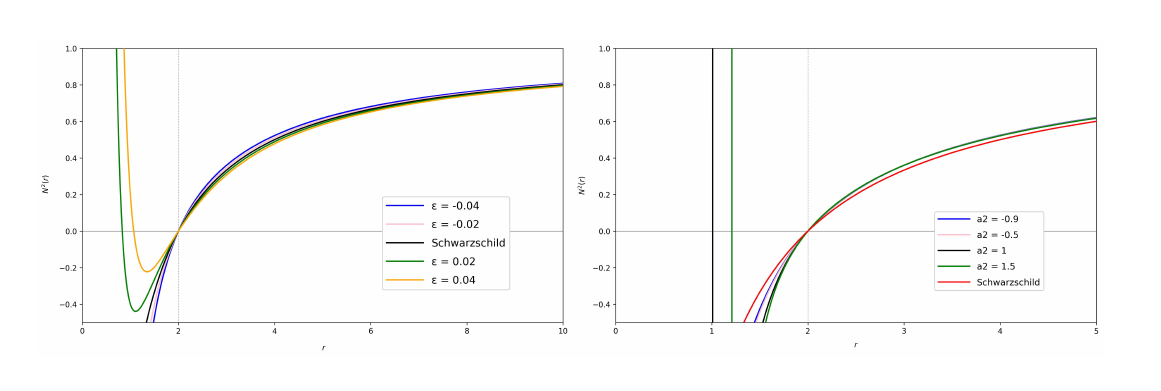}
\caption{The left panel displays metric functions for Schwarzschild (black) and modified Schwarzschild black holes with varying \(\varepsilon\) values (-0.04 blue, -0.02 pink, 0.02 green, 0.04 yellow), using \(a_2 = 1\). The right panel compares Schwarzschild (red) with modified black holes having different \(a_2\) values (-0.9 blue, -0.5 pink, 1 black, 1.5 green), where \(\varepsilon = -0.04\).}
\label{a}
\end{figure*}

Solving equation \eqref{8} yields the horizon conditions for different values of \(\varepsilon\) and \(a_2\), where \(r_0 = 2M\). For precise Schwarzschild black hole (SBH) shadow simulation, \(\varepsilon\) needs to remain minimal. In the following content, it is assumed that \(\varepsilon\) is within a reasonable range, i.e., \(\varepsilon \in [-0.04, 0.04]\)\cite{Xu:2024cfe}. As shown in Figure \ref{a}.

We find that when fixing \( a_2 = 1 \), the closer the value of \( \varepsilon \) is to 0, the closer it is to the Schwarzschild solution. When fixing \( \varepsilon = -0.04 \), the influence of different \( a_2 \) values on the metric function is difficult to distinguish. Therefore, in the following content, we take \( a_2 = 1 \). The variations of \( \varepsilon \) and \( a_2 \) show the metrics under two different scenarios.

\section{Motion Equations and Potential Function for Konoplya-Zhidenko Deformed Black Holes}\label{3.0}

Understanding photon behavior near Konoplya-Zhidenko deformed black holes requires deriving their motion equations and effective potential. We start from the Lagrangian and use conserved quantities and variable substitutions for step-by-step analysis.

\begin{figure}[]
\includegraphics[width=0.5\textwidth]{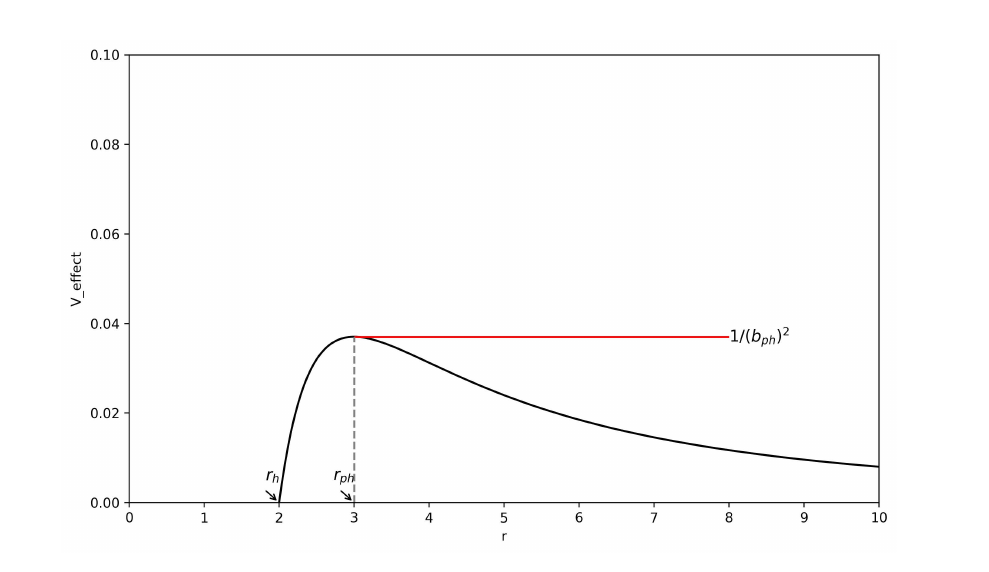}
\caption{
The graph showing the Schwarzschild black hole's effective potential \(V_{\text{eff}}(r)\) versus radius, and the relationship between \(r_{\text{ph}}\) and \(b_{\text{ph}}\).}
\label{b}
\end{figure}

For the given line element \eqref{1}, the corresponding Lagrangian is
\begin{equation}\label{9}
\mathcal{L} = \frac{1}{2} \left[ -N^2(r) \dot{t}^2 + \frac{B^2(r)}{N^2(r)} \dot{r}^2 + r^2 \dot{\theta}^2 + (r^2 \sin^2\theta) \dot{\phi}^2 \right].
\end{equation}
The dot denotes the \(\tau\)-derivative. Due to spherical symmetry and without loss of generality, the analysis simplifies the Lagrangian by confining it to the equatorial plane (\(\theta = \pi/2\), \(\dot{\theta} = 0\))
\begin{equation}\label{10}
\mathcal{L} = \frac{1}{2} \left[ -N^2(r) \dot{t}^2 + \frac{B^2(r)}{N^2(r)} \dot{r}^2 + r^2\dot{\phi}^2 \right].
\end{equation}
The corresponding canonical momentum is
\begin{equation}\label{11}
\begin{split}
p_t=\frac{\partial \mathcal{L}}{\partial \dot{t}}=-N^2(r)\dot{t},\\
p_\phi =-\frac{\partial \mathcal{L}}{\partial \dot{\phi }} =-r^2\dot{t},\\
p_r=-\frac{\partial \mathcal{L}}{\partial \dot{r}}=-\frac{B^2(r)}{N^2(r)}\dot{r},\\
p_\theta =-\frac{\partial \mathcal{L}}{\partial \dot{\theta }} =-r^2\dot{\theta }=0.  
\end{split}   
\end{equation}
Since the value of the Lagrangian density for null geodesics is zero, their integrals of motion can be obtained in the following manner
\begin{equation}\label{12}
\begin{split}
\frac{\mathrm{d} p_t}{\mathrm{d} \tau } =\frac{\partial \mathcal{L}}{\partial t}=0,\\
\frac{\mathrm{d} p_\phi }{\mathrm{d} \tau } =-\frac{\partial \mathcal{L}}{\partial \phi }=0. 
\end{split}   
\end{equation}
That is, we have
\begin{equation}\label{13}
p_t = -N^2(r) \frac{dt}{d\tau} = \text{const} = E,   
\end{equation}
and
\begin{equation}\label{14}
p_\phi = -r^2 \frac{d\phi}{d\tau} = \text{const} = L.  
\end{equation}
Here \(L\) represents angular momentum perpendicular to the invariant plane. The expressions \( \dot{t} = -E / N^2(r) \) and \( \dot{\phi} = -L / r^2 \) shown in equations \eqref{13} and \eqref{14} are derived from the conservation of the Lagrangian
\begin{equation}\label{15}
-\frac{E^2}{N^2(r)} + \frac{B^2(r)}{N^2(r)} \dot{r}^2 + \frac{L^2}{r^2} = 2\mathcal{L} = 0.  
\end{equation}
By representing \(r\) as a function of \(\phi\) rather than \(\tau\), \( \dot{r} = \frac{dr}{d\phi} \dot{\phi} \), the equation can be obtained as
\begin{equation}\label{16}
\left(\frac{dr}{d\phi}\right)^2 = \frac{E^2}{L^2} \frac{r^4}{B^2(r)} - r^2 \frac{N^2(r)}{B^2(r)}. 
\end{equation}
Let
\begin{equation}\label{17}
u = \frac{1}{r},
\end{equation}
then the fundamental equation of the problem is obtained \cite{Gan:2021xdl}
\begin{equation}\label{18}
B^2(\frac{1}{u} ) \left(\frac{du}{d\phi}\right)^2 = \frac{1}{b^2} - u^2 N^2(\frac{1}{u} ). 
\end{equation}
Where impact parameter \(b = L/E\) defines the null geodesic equation. We define the effective potential as
\begin{equation}\label{19}
V_{\text{eff}}(r) = u^2 N^2(\frac{1}{u} ).
\end{equation}
It is particularly worth noting that in the current context of discussion, we define circular null geodesics as photon spheres. The photon sphere coincides exactly with the effective potential \(V_{\text{eff}}(r)\) extremum point. In other words, the precondition for the photon sphere presence is \cite{Gan:2021pwu}
\begin{equation}\label{20}
V_{\text{eff}}(r_{ph} ) =\frac{1}{b^2_{ph}}, 
\end{equation}
and
\begin{equation}\label{21}
V_{\text{eff}}^\prime(r_{\text{ph}}) = 0.
\end{equation}
Here, \( r_{\text{ph}} \) indicates the photon sphere's radius, and \( b_{\text{ph}} \) represents the associated critical impact parameter. Their correlation is illustrated in Figure \ref{b}.

\begin{figure*}[]
\includegraphics[width=1\textwidth]{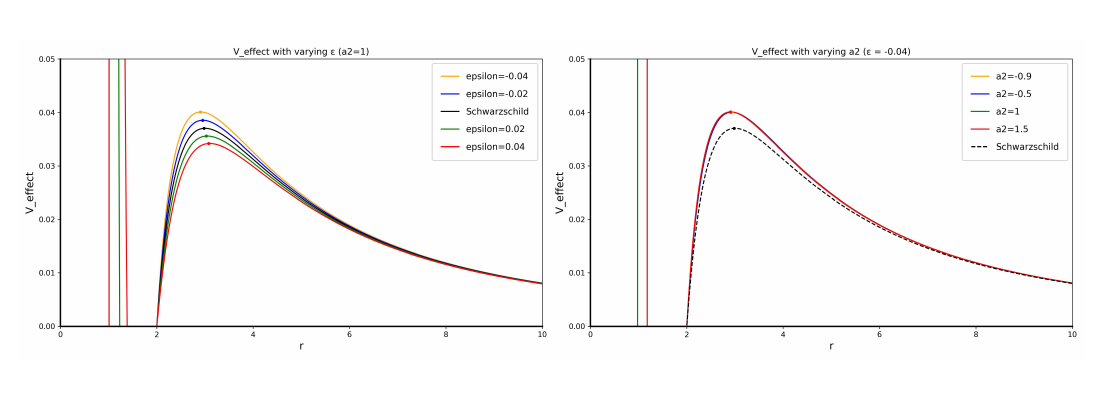}
\caption{
Left panel presents \(V_{\text{eff}}(r)\) against \(r\) with varying \(\varepsilon\) values at fixed \(a_2 = 1\), while right panel shows \(V_{\text{eff}}(r)\) against \(r\) with different \(a_2\) values at fixed \(\varepsilon = -0.04\).}
\label{c}
\end{figure*}

Fig. \ref{b} shows a SBH with \(r_h=2\) (EH radius) and \(r_{\text{ph}}=3\) (photon sphere radius). Figure \ref{c} displays effective potential \(V_{\text{eff}}(r)\) curves against radius when varying one parameter while keeping the other fixed.

From Figure \ref{c} shows that when \( a_2 \) is fixed, whether \( \varepsilon \) is positive or negative, its absolute value must be sufficiently small to well approximate the Schwarzschild solution. When \( \varepsilon \) is fixed, changes in different \( a_2 \) values show nearly identical effects on the physical quantities. This similarity in the effects of different \(a_2\) values further indicates that in the subsequent investigations, We'll maintain constant \(a_2\) to examine how various \(\varepsilon\) values affect the black hole.

The peak of the effective potential \( V_{\text{eff}} \) (corresponding to \( \delta V_{\text{eff}} < 0 \)) and the minimum point (corresponding to \( \delta V_{\text{eff}} > 0 \)) determine the radii of unstable photon orbits and bound photon motion, respectively. This difference in dynamic behavior originates from the fact that when photons in unstable orbits are perturbed, their trajectories change significantly—falling into the EH or escaping to the distant region—while photons in stable orbits maintain their state. Given that unstable photon orbits directly determine the optical characteristics of black hole accretion phenomena, particularly the radiation intensity distribution and the morphology of the shadow boundary received by distant observers \cite{Gan:2021pwu}, this paper will focus on analyzing the physical properties of such unstable photon orbits and their manifestations in astronomical observations.

\section{Multi-order Imaging Mechanism of Photon Rings in Strong-Field Spacetime}\label{4.0}

In this section, using the inverse ray-tracing technique, we conduct an in-depth analysis of the null geodesic characteristics of photon motion and the interaction mechanism between photons and a thin accretion disk in the general black hole spacetime background under the Konoplya-Zhidenko deformation rule. We model the accretion disk as an idealized thin structure, with observations made from the black hole's polar direction. The black hole's intense gravitational field strongly distorts light trajectories, enabling photons to traverse the accretion disk plane multiple times and form multiple overlapping orbiting trajectories. To this end, this study first systematically classifies photon trajectories based on the number of geodesic-disk intersections; Second, it quantifies the contribution weights of different photon trajectory paths to the total radiation intensity. By accurately calculating the cumulative brightness distribution of light rays under the spacetime deformation effects of the Konoplya-Zhidenko metric, this research develops the black hole's optical imaging properties without background illumination, emphasizing how multiple overlapping effects influence shadow and halo formations.
\subsection{Null Geodesic}\label{4.1}

According to null geodesic equation \eqref{18}, photon paths strongly correlate with impact parameter \(b\). Reference \cite{Gralla:2019xty} explains that for parallel null geodesics from the north pole direction, the impact parameter range depends on how many times these geodesics intersect the accretion disk, with the following relationship
\begin{equation}\label{22}
n=\frac{\phi }{2\pi }. 
\end{equation} 
In polar coordinates, \(\phi\) represents the total polar angle change along the null geodesic's path. For \(b < b_{\text{ph}}\), the total polar angle change outside the EH equalss \cite{Peng:2020wun,Yang:2022btw}
\begin{equation}\label{23}
\phi = \int_{0}^{u_h} \frac{1}{\sqrt{\frac{\left(\frac{1}{b^2} - N^2(\frac{1}{u} )u^2\right)}{B^2(\frac{1}{u} )}}}du.
\end{equation}  
Where \(u_h = \frac{1}{r_h}\) and \(r_h\) represents the EH radius. When \(b > b_{\text{ph}}\), the polar angle's total variation becomes 
\begin{equation}\label{24}
\phi = 2 \int_{0}^{u_{\text{max}}} \frac{1}{\sqrt{\frac{\left(\frac{1}{b^2} - N^2(\frac{1}{u})u^2\right)}{B^2(\frac{1}{u})}}} du.
\end{equation}  
Here, \(u_{\text{max}} = \frac{1}{r_{\text{min}}}\), with \(r_{\text{min}}\) being the null geodesic's closest approach to the black hole. When examining radiation near the black hole, observed intensity closely relates to how many times the geodesic crosses the accretion disk, which depends on b according to \cite{Peng:2020wun,Yang:2022btw}
\begin{equation}\label{25}
n(b)=\frac{2m-1}{4},m=1,2,3\cdots.  
\end{equation} 
Each value of \(m\) yields two solutions, \(b_m^\pm\), where \(b_m^-\) represents the smaller solution and \(b_m^+\) the larger one. Therefore, the intersection of null geodesics with the accretion disk can be classified into the following three types: The first type, a single intersection between null geodesic and accretion disk plane represents direct emission, occurring when \(n < \frac{3}{4}\). The second type, in the case of \(\frac{3}{4}<n<\frac{5}{4}\), the null geodesic intersects the disk plane twice, forming the lensing ring. For the third type with \(n > \frac{5}{4}\), null geodesics cross the disk plane \( \ge 3 \) times, forming photon ring emission.

\begin{figure}[]
\includegraphics[width=0.5\textwidth]{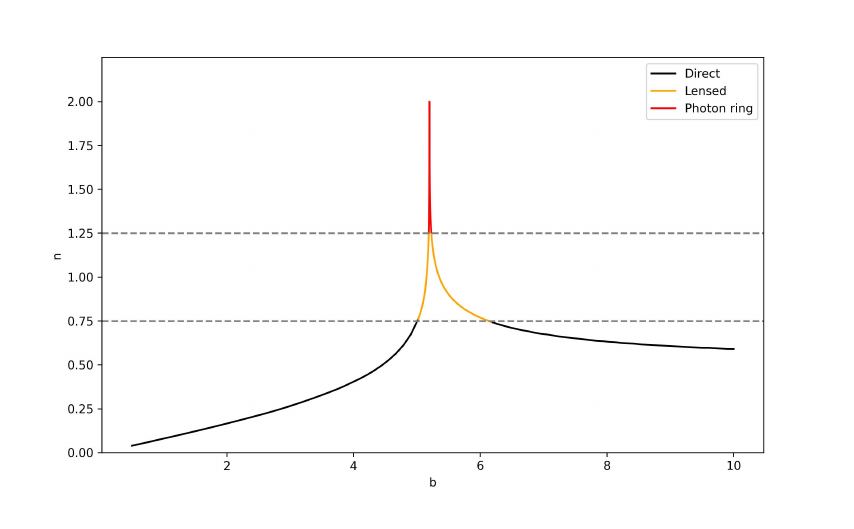}
\caption{
For a Schwarzschild black hole, the plot shows how \(n\) varies with impact parameter \(b\). The black curve is for direct emission, the orange one for lens ring emission, and the red one for photon ring emission. }
\label{d}
\end{figure}

Fig. \ref{d} shows the \(n-b\) curve for a SBH.
 
As illustrated in Figure \ref{d}, these light rays can be categorized as follows:

(a)Direct emission: \( n < \frac{3}{4} \leftrightarrow  b \in (0,\, b_2^-) \cup (b_2^+,\, \infty) \);

(b)Lensing ring : \( \frac{3}{4} < n < \frac{5}{4} \leftrightarrow  b \in (b_2^-,\, b_3^-) \cup (b_3^+,\, b_2^+) \);

(c)Photon ring : \( n > \frac{5}{4} \leftrightarrow  b \in (b_3^-,\, b_3^+) \).

\begin{figure*}[]
\includegraphics[width=1\textwidth]{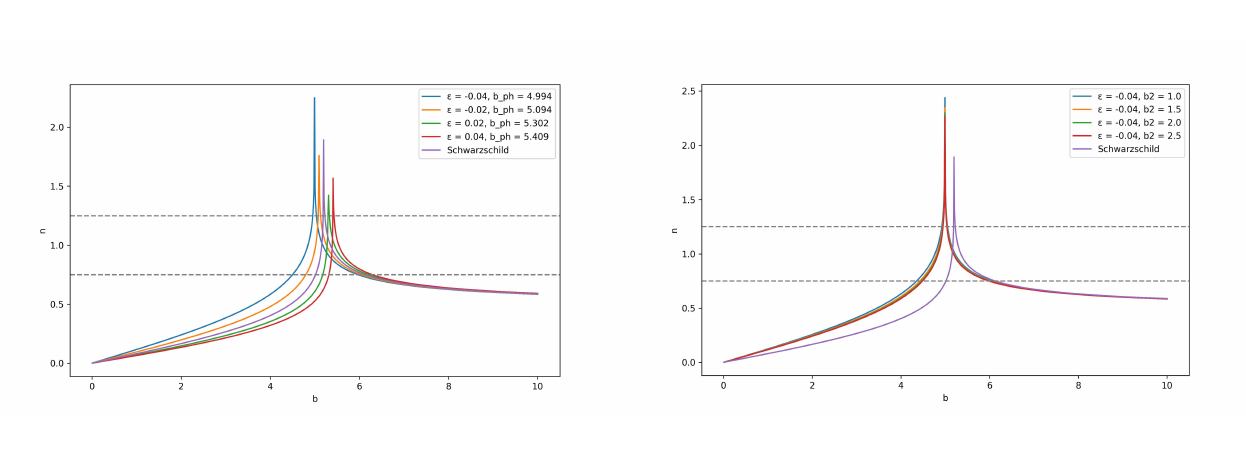}
\caption{
The curve illustrating the  correlation between \(n\) and the collision parameter \(b\). In the left figure, \(a_2 = 1\) and \(b_2 = 3\), and there are five cases where \(\varepsilon\) takes the values of -0.04 (blue), -0.02 (orange), 0.02 (green), 0.04 (red) respectively, as well as the Schwarzschild solution (purple). In the right figure, \(a_2 = 1\) and \(\varepsilon=-0.04\), and there are five cases where \(b_2\) takes the values of 1 (blue), 1.5 (orange), 2 (green), 2.5 (red) respectively, as well as the Schwarzschild solution (purple).}
\label{e}
\end{figure*}

Similarly, by using the null geodesic equation (Equation \eqref{18}), the \(n-b\) correlation curve can be derived, as shown in Figure \ref{e}. The left figure shows the n-b curve with fixed \(b_2, a_2\) and varying \(\varepsilon\). The right figure shows the n-b curve with fixed \(a_2, \varepsilon\) and varying \(b_2\). Meanwhile, the Schwarzschild case is also presented.

As can be seen from Figure \ref{e}, in the left figure, under the condition that \(a_2\) and \(b_2\) are fixed, the following general pattern emerges: When the impact parameter \(b\) is small, the corresponding values of \(n\) for each curve are consistently small, with gradual growth. As the impact parameter \(b\) grows, the deflection of the null geodesic first intensifies and then diminishes. Moreover, when the value of \(\varepsilon\) approaches 0, the value of \(b_{ph}\) gradually approaches \(3\sqrt{3}\) for the SBH. Simultaneously, the impact parameter ranges for the lensing and photon rings gradually narrow. In contrast, the right figure shows the changes of the curves under different values of \(b_2\). It can be seen from it that the change trends among the curves are nearly the same, making it difficult to effectively distinguish the cases of different values of \(b_2\) based on these curves. Based on the above analysis, in order to ensure the accuracy and effectiveness of subsequent research, the value of \(b_2\) will be fixed at 3 during the subsequent research process. 

\begin{figure*}[]
\includegraphics[width=1\textwidth]{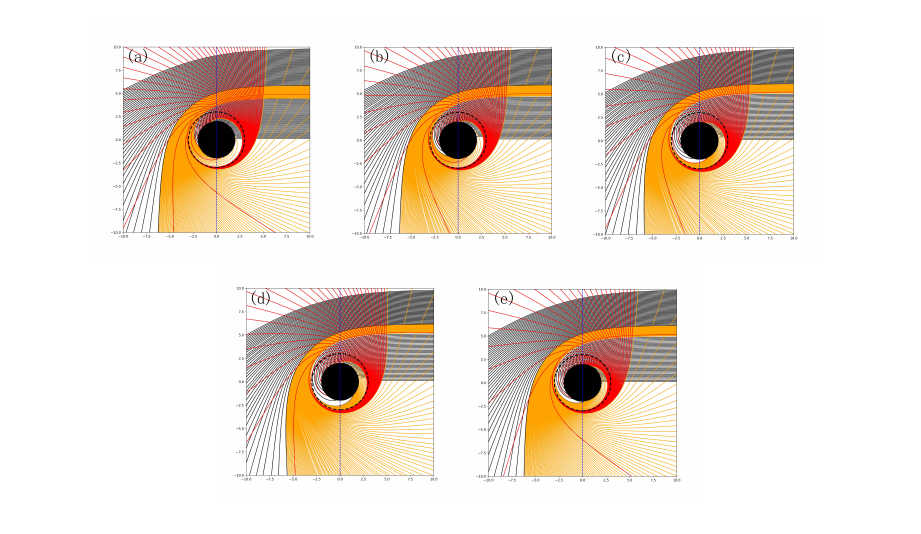}
\caption{
For black holes, photon trajectories are shown for \(\varepsilon = -0.04\) (a), \(\varepsilon = -0.02\) (b), \(\varepsilon = 0.02\) (c), \(\varepsilon = 0.04\) (d), and the Schwarzschild scenario (e). The black dashed circle is the photon sphere, while the blue vertical dashed line marks the disk plane's cross - section. Null geodesics are color - coded by the number of intersections with the accretion disk: black for one intersection, orange for two, and red for three or more. In all these figures, \(a_2 = 1\) and \(b_2 = 3\).}
\label{f}
\end{figure*}

Next, use the null geodesic equation (Equation \eqref{18}) to draw the null geodesic diagram, as shown in Figure \ref{f}. 

As depicted in Figure \ref{f},

(a)The impact parameter \(b\) in the figure carries a clear physical meaning: it corresponds to the impact parameter (defined as the asymptotic distance of the light ray from the black hole's central axis) when the null geodesic extends toward infinity (North Pole direction). Observationally, \(b\) characterizes the angular momentum-to-energy ratio of the photon, serving as a key parameter for studying light propagation and strong gravitational field effects around black holes.

(b)Theoretical analysis and numerical computations show the EH radius \(r_h\) remains equal to the SBH value. Meanwhile, the photon sphere radius \(r_{\text{ph}}\) and the critical impact parameter \(b_{\text{ph}}\) increase monotonically with \(\varepsilon\).

(c)Analysis of the influence of \(\varepsilon\) reveals that general black holes exhibit qualitative similarities with SBHs. As \(\varepsilon\) increases from \(-0.04\) to zero, The radial distance of the photon sphere, corresponding to the geodesic with maximum angular deflection, rises from a value below the Schwarzschild limit to the Schwarzschild limit of \(3\sqrt{3}\); The reverse trend occurs as \(\varepsilon\) decreases from \(0.04\) to zero. This demonstrates the critical role of \(\varepsilon\) in modulating spacetime geometry and light trajectories.

(d)When the collision parameter \(b\) rises, the curvature extent of the null geodesic first goes up and then drops, reaching a relative maximum at \(b = b_{ph}\). The null geodesics are color-coded according to their intersections with the accretion disk, with colors transitioning in the sequence transitioning from black to orange, then to red, followed by orange, and returning to black. intuitively presenting the change in their degree of bending, which provides a basis for exploring the light propagation mechanism around the black hole.

More precise calculations and analysis can determine the boundary values of collision parameter \(b\) for varying null geodesic-accretion disk intersection numbers. The pertinent computational results are compiled in Table \ref{table1}. This table lists boundary values of collision parameter b for different intersection cases and computes \(r_h\) (EH radius), \(r_{\text{ph}}\) (photon sphere radius), \(b_{\text{ph}}\) (critical impact parameter), and \(r_{\text{isco}}\) (ISCO radius). The radius \(r_{isco}\) can be determined using Equation \eqref{26} \cite{Rezzolla:2014mua}
\begin{equation}\label{26}
r_{isco}=\frac{3N(r_{isco})(N(r_{isco}))'}{3{(N(r_{isco}))'^{2}}-N(r_{isco})(N(r_{isco}))'' }  
\end{equation}
Here, the dot denotes the derivative with respect to radial distance r.          
\begin{table*}[]
\centering
\begin{tabular}{p{3.2cm}p{1.7cm}p{1.7cm}p{1.7cm}p{1.7cm}p{1.7cm}p{1.7cm}p{1.7cm}p{1.7cm}}
\hline\hline
\rule{0pt}{12pt}$\varepsilon$ & $b_{2}^{-}$ & $b_{2}^{+}$ & $b_{3}^{-}$ & $b_{3}^{+}$ & $r_{h} $ & $r_{ph}$ & $b_{ph}$ & $r_{isco}$  \\
\hline
\rule{0pt}{12pt} 
$-0.04$ &4.510 & 5.960 & 4.988 & 5.025 & 2.000 & 2.904 & 4.994 & 5.727  \\
\rule{0pt}{12pt} 
$-0.02$ & 4.805 & 6.036 & 5.085 & 5.125 & 2.000 & 2.953 & 5.094 & 5.862 \\
\rule{0pt}{12pt} 
$Schwarzschild$ & 5.020 & 6.170 & 5.180 & 5.230 & 2.000 & 3.000 & $3\sqrt{3}$ & 6.000 \\
\rule{0pt}{12pt} 
$0.02$ & 5.180 & 6.200 & 5.290 & 5.333 & 2.000 & 3.028 & 5.302 & 6.140  \\
\rule{0pt}{12pt} 
$0.04$ & 5.320 & 6.280 & 5.400 & 5.438 & 2.000 & 3.077 & 5.409 & 6.282 \\
\hline\hline
\end{tabular}
\caption{With \(a_2 = 1\) and \(b_2 = 3\) fixed, boundary values of impact parameter b are given for five \(\varepsilon\) cases: -0.04, -0.02, 0.02, 0.04, and the Schwarzschild solution. Corresponding values of \(r_h\) (EH radius), \(r_{\text{ph}}\) (photon sphere radius), \(b_{\text{ph}}\) (critical impact parameter), and \(r_{\text{isco}}\) (innermost stable orbit) are also listed. }
\label{table1}
\end{table*}
\subsection{Revise the existing constraints on $\varepsilon$}\label{4.2}

According to  the findings of the EHT collaboration regarding the photon rings of M87* and Sgr A* at the Galactic Center, as reported in \cite{EventHorizonTelescope:2019dse, EventHorizonTelescope:2022wok}, the primary parameters are as follows:

(a)For M87*: The shadow's angular diameter is $\theta_{M87^*} = (42 \pm 3)\ \mu$as, the distance from Earth is $D_{M87^*} = 16.8^{+0.8}_{-0.7}\ \text{Mpc}$, and the mass is $M_{M87^*} = (6.5 \pm 0.9) \times 10^{9}\ M_{\odot}$.
  
(b)For Sgr A*: The shadow's angular diameter is $\theta_{Sgr A^*} = (48.7 \pm 7)\ \mu$as, the distance from Earth is $D_{Sgr A^*} = (8277 \pm 33)\ \text{pc}$, and the mass is $M_{Sgr A^*} = (4.3 \pm 0.013) \times 10^{6}\ M_{\odot}$.

Using calculations derived from general relativity, the physical diameters of the shadows for M87* and Sgr A* are determined to be $d_{sh}^{M87^*} = (11 \pm 1.5)$ and $d_{sh}^{Sgr A^*} = (9.5 \pm 1.4)$, respectively \cite{Luo:2024avl}.

\begin{figure*}[]
\includegraphics[width=1\textwidth]{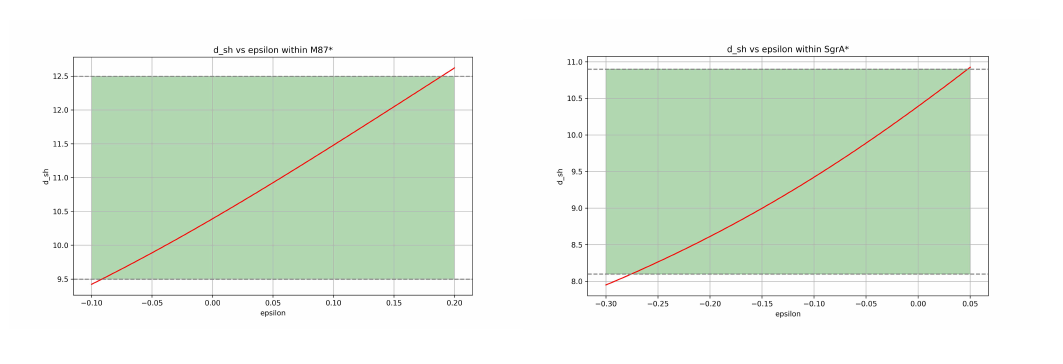}
\caption{
The EHT collaboration constrains the parameter \(\varepsilon\) in this metric by analyzing the shadow diameters of the M87* and Sgr A* black holes. The red curve illustrates the variation in black hole shadow diameter as \(\varepsilon\) changes, while the green highlighted area represents the span of observed shadow diameter values, thus determining the value of \(\varepsilon\).}
\label{g}
\end{figure*}

For a non-moving observer positioned at \(r_0\), the theoretical expression for the black hole shadow's radius can be derived \cite{EventHorizonTelescope:2020qrl}
\begin{equation}\label{27}
r_{sh}=\frac{r_{ph}}{\sqrt{-g_{tt}(r_{ph})}}.  
\end{equation}
From Equation \eqref{1}, we can obtain \(r_{sh}=\frac{r_{ph}}{\sqrt{N^2(r_{ph})}}\). According to the spatial symmetry, the diameter of the shadow satisfies \(d_{sh} = 2r_{sh}\). As shown in Figure \ref{g}, the shadow diameter measurements from the EHT impose stringent observational constraints on the coupling parameter \(\varepsilon\) in the theoretical model presented in this paper.

By analyzing Figure \ref{g}, it can be seen that, under the confidence level based on $d_{sh}^{M87^*}$, the parameter \(\varepsilon\) is constrained within the interval \(-0.09\lesssim\varepsilon\lesssim0.19\); under the confidence level of \(d_{sh}^{Sgr A^*}\), the parameter \(\varepsilon\) is constrained within the interval \(-0.280\lesssim\varepsilon\lesssim0.047\). Obviously, the upper limit constraint of the parameter \(\varepsilon\) obtained from Sgr A* is stricter than that from M87*, but the lower limit constraint of the parameter \(\varepsilon\) obtained from M87* is stricter than that from Sgr A*. In addition, for both M87* and Sgr A*, the parameter \(\varepsilon\) ranges from \(-0.04\) to \(0.04\). Therefore, in order to simplify the problem, our subsequent discussions about \(\varepsilon\) will be limited to this range. Moreover, regardless of whether \(\varepsilon\) approaches 0 from the positive or negative direction, the black hole's characteristics approach SBH properties. This highlights the regulatory influence of \(\varepsilon\) on black hole properties, offering a foundation for investigating black hole behavior.
\subsection{Black Hole Appearance with Konoplya-Zhidenko Modifications}\label{4.3}
 
We'll now simulate black holes with varying \(\varepsilon\) values to show their photon rings and shadows from a distant viewpoint. For this research, we consider light to come only from the accretion disk. According to the theories of optics and the light rays in the gravitational field, Light rays from the accretion disk follow null geodesic paths. Since the null geodesic extracts energy each time it intersects, the more times it intersects, the higher the energy and brightness of the light rays will be, thus significantly altering the observed intensity distribution.

The black hole is encircled by a flat, stationary disk of matter. As the matter in the disk radiates evenly in all directions, and the radiation frequency received by the observer is $\upsilon_e$. In the case of monochromatic light, using Liouville's theorem, we can find the intensity that reaches the observer \cite{Gan:2021xdl,Wang:2023vcv}
\begin{equation}\label{28}
I_o(r,\nu_0) = g^3I_e(r,\nu_e).  
\end{equation}
The redshift factor \(g = \frac{\nu_0}{\nu_e} = \sqrt{N^2(r)}\), where \(\nu_0\) and \(\nu_e\) are observed and emitted light frequencies, respectively. Additionally, \(I_o(r,\nu_0)\) and \(I_e(r,\nu_e)\) denote the specific intensities of monochromatic light observed and emitted at radius \(r\), respectively. By integrating the frequencies of all observed \(I_o(r,\nu_0)\), the total observed intensity across the entire wavelength band can be determined
\begin{equation}\label{29}
\begin{split}
I_{\text{obs}}(r) &= \int I_o(r,\nu_0) \, d\nu_0 \\
&= \int g^4 I_e(r,\nu_e) \, d\nu_e \\
&= \left(N^2(r)\right)^2 I_{\text{em}}(r).
\end{split}
\end{equation}
Owing to the inclusion of the transfer function \( r_m(b) \), this equation is now reformulated as a function of \(b\) rather than \(r\). Here, \(I_{em}(r)=\int I_e(r,\nu_e)d\nu_e\)gives the overall radiation strength. Light rays reaching the observer gain energy at points where they cross the accretion disk, thereby increasing its brightness. Therefore, based on the ray's closest approach distance and radiation properties, the light ray may acquire energy once, twice, or more times, thereby enhancing its brightness. So, complete brightness results from combining each crossing's contribution \cite{Gralla:2019xty,Peng:2020wun}, that is
\begin{equation}\label{30}
I_{obs}(b)=\sum_{m} \left(N^2(r))^2 I_{em}(r)\right|_{r = r_m(b)}. 
\end{equation}
\begin{figure}[]
\includegraphics[width=0.5\textwidth]{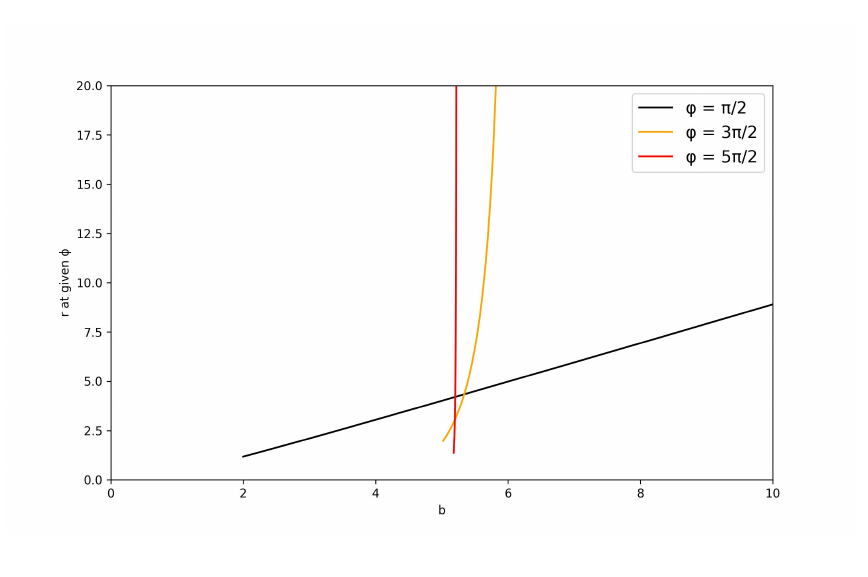}
\caption{
For the Schwarzschild black hole, we show three transfer functions: first (black), second (orange), and third (red).}
\label{h}
\end{figure}
\begin{figure}[]
\includegraphics[width=0.5\textwidth]{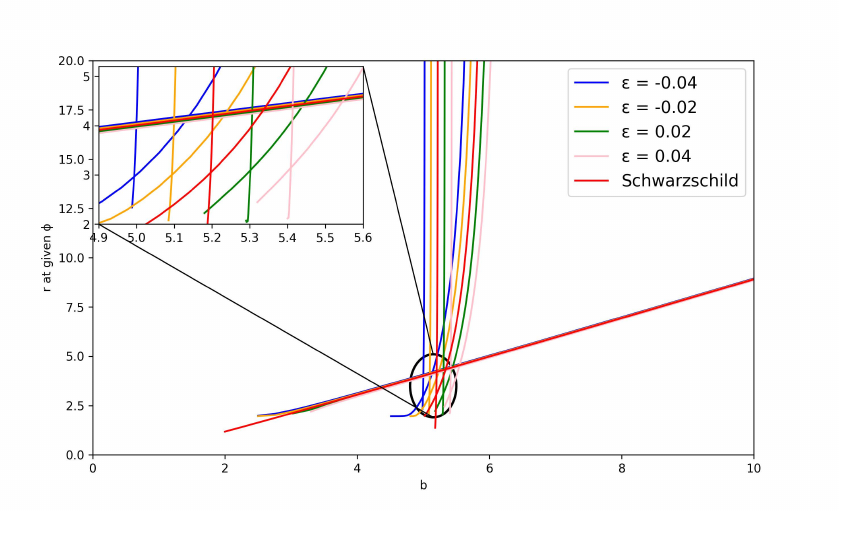}
\caption{
The first three transfer functions in the cases where \(\varepsilon\) takes values of -0.04 (blue), -0.02 (orange), 0.02 (green), 0.04 (pink) respectively, and for the Schwarzschild model (red).}
\label{i}
\end{figure}
\begin{figure*}[]
\includegraphics[width=1\textwidth]{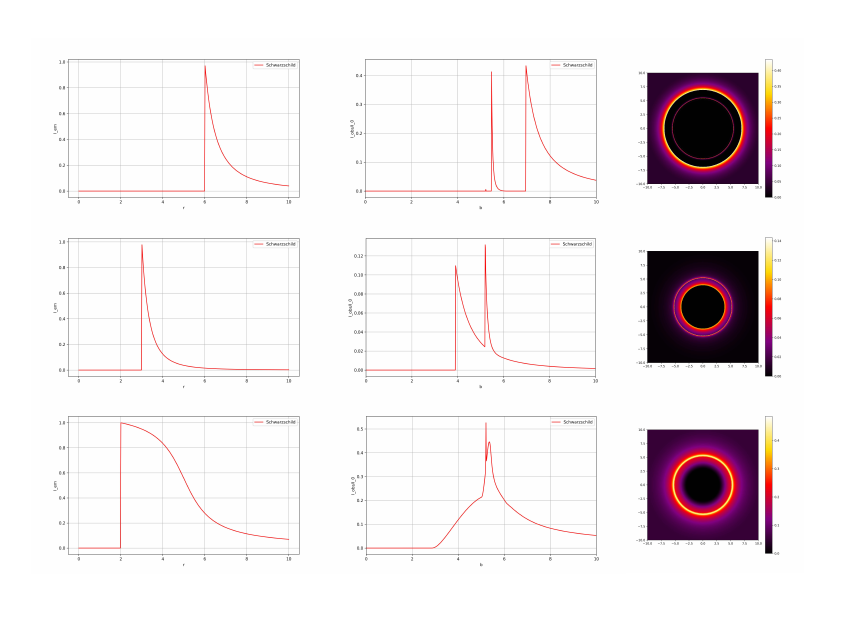}
\caption{
It displays emission intensity vs. radius for Schwarzschild black holes using three models (left), total observed intensity vs. impact parameter (middle), and the visual appearance (right).}
\label{j}
\end{figure*} 
\begin{figure*}[]
\includegraphics[width=1\textwidth]{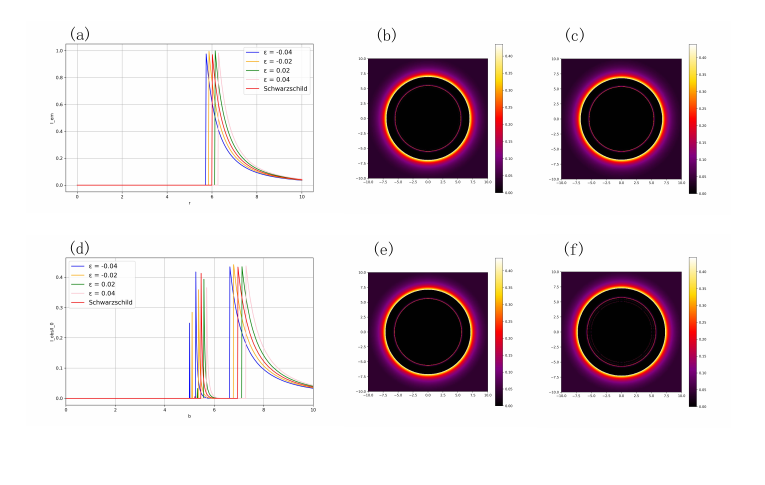}
\caption{
For the first model, the emission intensity curve as a function of radius \(r\) (a), the total observed intensity curve with respect to impact parameter \(b\) (d), and the optical appearances viewed by the observer at \(\varepsilon = -0.04\) (b), \(\varepsilon = -0.02\) (c), \(\varepsilon = 0.02\) (e), and \(\varepsilon = 0.04\) (f) are presented.}
\label{k}
\end{figure*}
\begin{figure*}[]
\includegraphics[width=1\textwidth]{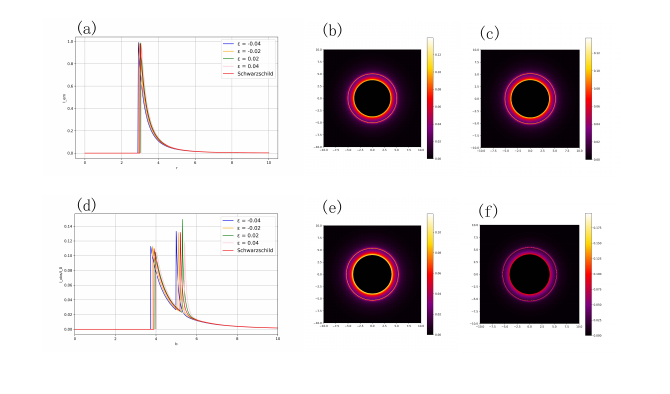}
\caption{
In the second - model context, there's the curve of emission intensity as a function of radius \(r\) (a), the curve of total observed intensity in relation to impact parameter \(b\) (d), and the optical appearances perceived by the observer at \(\varepsilon = - 0.04\) (b), \(\varepsilon = - 0.02\) (c), \(\varepsilon = 0.02\) (e), and \(\varepsilon = 0.04\) (f). }
\label{l}
\end{figure*}
\begin{figure*}[htbp]
\includegraphics[width=1\textwidth]{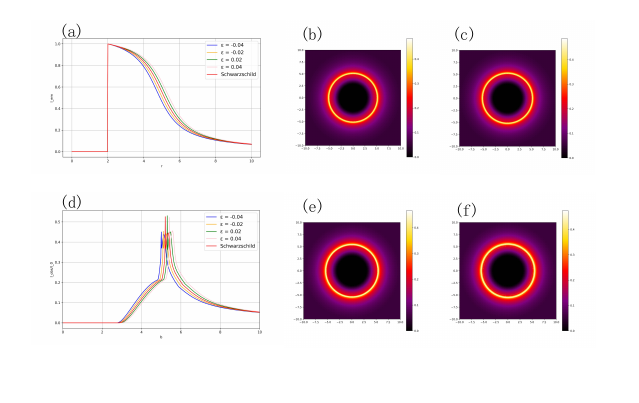}
\caption{
For the third model, there is the curve depicting the variation of emission intensity with radius \(r\) (a), the curve showing how the total observed intensity changes with impact parameter \(b\) (d), along with the optical appearances observed by the viewer at \(\varepsilon = - 0.04\) (b), \(\varepsilon = - 0.02\) (c), \(\varepsilon = 0.02\) (e), and \(\varepsilon = 0.04\) (f). }
\label{m}
\end{figure*}
The function \(r_m(b)\) (\(m = 1, 2, 3, \ldots\)), called the transfer function, maps the impact parameter b of a light ray to the radial coordinate where it intersects the accretion disk at the \(m\)th time. In addition, according to Reference \cite{Gralla:2019xty}, The slope \(dr/db\) indicates the demagnification factor for each \(b\). For the convenience of analysis, our model excludes light absorption and reflection by the disk. For the case of \(m > 3\), photon rings' contribution to total brightness is negligible. The first three transfer functions \(r_m(b)\) are formulated as described in \cite{Yang:2022btw}
\begin{equation}\label{31}
r_1(b)=\frac{1}{u(\frac{\pi}{2},b)}, if b_1^- < b < +\infty,
\end{equation}
\begin{equation}\label{32}
r_2(b)=\frac{1}{u(\frac{3\pi}{2},b)}, if b_2^- < b < b_2^+,
\end{equation}
\begin{equation}\label{33}
r_3(b)=\frac{1}{u(\frac{5\pi}{2},b)}, if b_3^- < b < b_3^+. 
\end{equation}
In this context, \(u(\phi, b)\) satisfies Equation \eqref{18}. The graphical representation in Figure \ref{h} depicts the first three Schwarzschild transfer functions \(r_m(b)\). 

Figure \ref{h} analysis suggests that direct radiation, the lensing ring, and the photon ring may all contribute to the composition of the first transfer function; Regarding the second transfer function, it emerges from both the lensing ring and photon ring components, whereas the third transfer function is exclusively derived from the photon ring. Figure \ref{i} displays the first three \(r_m(b)\)) transfer functions across various  \(\varepsilon\)values.

As shown in Figure \ref{i}, when \(\varepsilon\) approaches zero from either the negative or positive direction, there is significant intersection among the three transfer function regions, with all gradually trending toward alignment with the Schwarzschild model. Across different parameter settings and physical scenarios, the incline of the transfer function increases significantly with the value of \(m\). The steeper gradient suggests a reduced impact on the overall light intensity, which aligns with theoretical predictions and further validates the model's applicability in this context. Additionally, it is clearly observable that the slopes of the transfer function curves corresponding to all \(r_1(b)\) are approximately unity, this suggests that the radial position \(r_1\), where the null geodesic initially crosses the accretion disk, demonstrates a nearly linear relationship with the impact parameter \(b\).

To validate the forecast of the transfer function, a specific radiation profile of the accretion disk will be considered, and the total observed intensity of each radiation will be calculated by using formula \eqref{30}. We expect to understand the contributions of the radiations from the direct radiation, The gravitational lensing ring and photon ring each affect the black hole's total luminosity and appearance. In this study, we apply three uncomplicated radiation function models \cite{Li:2021riw}. 
The inaugural emission model adopts a function with decay following a squared power relationship
\begin{equation}\label{34}
I_{em}(r)=\begin{cases}I_0\left[\frac{1}{r-(r_{isco}-1)}\right]^2, & r > r_{isco}\\0, & r\leq r_{isco}\end{cases}, 
\end{equation}
for the second emission framework, we adopt a function characterized by cubic attenuation
\begin{equation}\label{35}
I_{em}(r)=\begin{cases}I_0\left[\frac{1}{r-(r_{ph}-1)}\right]^3, & r > r_{ph}\\0, & r\leq r_{ph}\end{cases},
\end{equation}
the third emission model is a function with a slower decay
\begin{equation}\label{36}
I_{em}(r)=\begin{cases}I_0\frac{\frac{\pi}{2}-\tan^{-1}\left[r-(r_{isco}-1)\right]}{\frac{\pi}{2}-\tan^{-1} \left[r_{h}-(r_{isco}-1)\right]}, & r > r_{h}\\0, & r\leq r_{h}\end{cases}. 
\end{equation}
Among them, \(r_{h}\) is the EH radius, \(b_{ph}\) is the critical impact parameter, \(r_{ph}\) is the photon sphere radius, and \(r_{isco}\) is the ISCO radius. Their corresponding values can be shown in Table \ref{table1}. And the emission intensities reach their peaks at \(r_{isco}\), \(r_{ph}\) and \(r_{h}\) respectively. In the first two cases, the intensities decay sharply, while in the last case, the decay is slow. Here, \(I_0\) is the maximum intensity.

Incorporate the emission intensity and transfer functions into Equation \eqref{30}, Figure \ref{j} illustrates the resultant thin disk appearance near a SBH. For the SBH solution, emission and aggregate observed intensities are shown in the left and middle panels, respectively.

Figure \ref{j} presents three aspects of the SBH soln. Left column shows emission intensity curves as a function of radius across three models. The middle column depicts total observed intensity versus impact parameter, where the first model shows three distinct peaks from photon, lensing, and direct ring radiation. Photon ring peak is nearly imperceptible, contributing negligibly compared to the dominant lensing and direct radiation peaks. The right column illustrates the optical appearance from an external observer's perspective.

Next, We study the thin disk's appearance near a black hole under four \(\varepsilon\) profiles: -0.04, -0.02, 0.02, and 0.04. Figures \ref{k}, \ref{l} and \ref{m} respectively, for the first, second and third models, displays emission intensity versus radius \(r\), total observed intensity versus impact parameter \(b\), and the resulting optical appearance as seen by observers.  

Through the analysis of Figures \ref{k}, \ref{l} and \ref{m}, it can be known that:

(a)In the three toy models, the emission intensity all exhibits specific variation patterns. Specifically, the emission intensity peaks at \(r_{\text{isco}}\) (innermost stable orbit), \(r_{\text{ph}}\) (photon sphere), and \(r_h\) (event horizon), respectively. After that, for the first and second models, the emission intensity decays rapidly; While in the third model, the emission intensity shows a slow downward trend. As $\varepsilon$ approaches 0, the peak positions and amplitudes of the emission intensity under each model show a tendency to approach the situation of the Schwarzschild solution.           

(b)In the first model, the total observed intensity versus impact parameter \(b\) shows three distinct peaks representing photon ring, lensing ring, and direct ring radiation signatures. Notably, the photon ring occupies an extremely narrow region in the intensity profile. As the parameter \(\varepsilon\) approaches 0, peak locations and amplitudes of observed intensity converge to the Schwarzschild solution's behavior, with the first peak (associated with the photon ring) gradually weakening until it nearly vanishes. Its impact on the overall observed intensity is insignificant and can be disregarded. In contrast, the second and third models' intensity curves display only two prominent peaks from left to right, indicating that the total observed intensity is primarily dominated by these two features. When projecting total observed intensity onto a 2D plane to simulate the observer's view, the results manifest as two bright rings—a characteristic intuitively visualized in figures (b), (c), (e), and (f).

(c)Analysis of the optical appearance diagrams in the three toy models reveals that the first model exhibits a slender bright ring in the dark zone, identified as the lensing ring. Inside this lensing ring lies a smaller photon ring, which remains hidden due to its extremely low brightness and is thus highly challenging to capture observationally. These two rings contribute minimally to overall brightness, while the direct ring positions itself outside the lensing ring, where the brightness abruptly increases to an extremely high level. However, moving further outward from the direct ring, the brightness gradually decreases. In sharp contrast to the first model, the second model features a brighter direct ring situated inside the lensing ring. The third model, unlike the first two, displays only a single ring structure, where brightness transitions from dark to bright then back to dark.

In conclusion, within the research framework of the three models, changes in $\varepsilon$ significantly affect both the total observed intensity curve and the black hole image. Firstly, when $\varepsilon$ gradually approaches 0 from -0.04, the total observed intensity peak shifts to larger b values. In this process, as $\varepsilon$ gets closer to 0 and takes a value larger than -0.04, a photon ring with a larger radius can be observed in the presented black hole image. Conversely, when $\varepsilon$ gradually approaches 0 from 0.04, the total observed intensity curve's peak shifts toward smaller \(b\) values. Similarly, during the process of $\varepsilon$ approaching 0 from 0.04, when $\varepsilon$ is close to 0 and takes a value larger than 0, a photon ring with a larger radius will also appear in the black hole image. According to the above variation trends, for a general black hole under the Konoplya-Zhidenko deformation rule, its photon ring image shows no degeneracy. This characteristic provides a new feasible approach for constraining the extended theory of gravity.

Secondly, compared to the first model, the second model's total observed intensity shows a downward trend, which manifests optically as a relatively darker visual effect. In contrast to the first two models, the third model displays only a single-ring appearance, yet its total observed intensity is significantly higher. Theoretically, the third model ought to feature a two-ring structure. However, based on the observed intensity distribution, the peaks are too close together for the double-ring phenomenon to be clearly distinguished during observation. 

Finally, through meticulous investigations of different models, the second emission model shows the photon ring morphology of a generic black hole under the Konoplya-Zhidenko deformation rule is highly sensitive. This stems from the fact that the photon sphere radius \(r_{ph}\) of this black hole varies sensitively with the parameter \(\varepsilon\). This finding offers a new perspective and critical clues for advancing research on black hole physical properties and related gravitational theories, while enhancing our understanding of how matter radiation interacts with spacetime geometry around black holes.
                
\section{Conclusion}\label{5.0}

The paper explores optical characteristics of black holes under Konoplya-Zhidenko deformation by simulating a generalized Schwarzschild black hole. The study relies on the light intensities emitted by three toy models and discusses the influences of the deformation parameters \(\varepsilon\), \(a_2\), and \(b_2\) on its morphology.

Firstly, the impacts of the three parameters \(\varepsilon\), \(a_2\), and \(b_2\) on the event horizon were analyzed. The research findings indicate that to achieve a high-precision simulation of the Schwarzschild black hole's shadow, \(\varepsilon\) must be confined within an extremely narrow range. In this study, the value range of \(\varepsilon\) was set as \([-0.04, 0.04]\). To maintain positive metric coefficients beyond the event horizon and prevent naked singularities, modification parameters require \(a_2 > - 1\) and \(b_2 > - 1\). By keeping the value of \(\varepsilon\) constant and thoroughly exploring the influence of \(a_2\) on the metric function and the effective potential, it was discovered that when \(a_2\) meets the condition of \(a_2 > - 1\), the impacts of different values of \(a_2\) on the two aspects do not exhibit significant discrepancies. Consequently, in the subsequent research of this paper, \(a_2\) was fixed at 1. Likewise, different values of the deformation parameter \(b_2\) also show no pronounced differences in their effects on the research content. Thus, in this study, \(b_2\) was set to a fixed value of 3. According to the cosmic censorship hypothesis forbidding naked singularities, this paper thoroughly examines the optical features of general black holes under Konoplya-Zhidenko deformation, with the parameter conditions of \(\varepsilon\in [-0.04, 0.04]\), \(a_2 = 1\), and \(b_2 = 3\).

Subsequently, from the Lagrangian, we derived this black hole's geodesic equation and effective potential. The exact radius of the unstable photon sphere orbit is determined by the effective potential. Back-propagation ray-tracing revealed that parameter \(\varepsilon\) affects both the distribution and typology of null geodesics around this black hole. As \(\varepsilon\) gradually increases, the event horizon \(r_h\) remains unchanged (consistently equal to that of the Schwarzschild black hole), while The photon sphere radius \(r_{ph}\) , ISCO radius \(r_{isco}\), and critical impact parameter \(b_{ph}\) all increase. Concurrently, the impact parameter range for the lens and photon rings progressively narrows, and this change further affects the morphology of the black hole's photon ring. Subsequently, by integrating EHT observational data on the shadow diameters of \(M87^*\) and \(Sgr A^*\) black holes, we re-constrained the value range of the parameter \(\varepsilon\). Specifically, at the confidence level based on \(d_{\text{sh}}^{(M87^*)}\), \(\varepsilon\) is constrained within \(-0.09 \lesssim \varepsilon \lesssim 0.19\); For \(d_{\text{sh}}^{(Sgr A^*)}\), the constraint is \(-0.280 \lesssim \varepsilon \lesssim 0.047\). Notably, both confidence intervals highly overlap with the interval \([-0.04, 0.04]\). When \(\varepsilon\) approaches 0, this black hole's properties converge toward those of a Schwarzschild black hole.

Finally, We examine how thin accretion disks illuminate optical features of black hole under Konoplya-Zhidenko deformation, all three toy models show two distinct intensity peaks, creating two bright rings in the 2D image. Further analysis shows that as $\varepsilon$ increases, the peaks of the total observed intensity shift toward larger impact parameter \(b\), leading to an increase in the ring radius, with the photon ring radius \(r_{ph}\) being smaller than the Schwarzschild value when $\varepsilon < 0$ and larger than the Schwarzschild value when $\varepsilon > 0$, such that the closer $\varepsilon$ is to 0, the more closely the system approximates the Schwarzschild spacetime. In conclusion, black hole photon spheres and their images probe spacetime properties, with theoretical potential to distinguish the spacetime metrics of general black holes under different Konoplya-Zhidenko deformation rules and provide potential research directions and an important theoretical basis for constraining extended gravitational theories through upcoming black hole photon ring observations.
\section*{Acknowledgements}

This work was supported by Guizhou Provincial Basic Research Program (Natural Science)(Grant No.QianKeHeJiChu[2024]Young166), the Special Natural ScienceFundofGuizhouUniversity(GrantNo.X2022133),the National Natural Science Foundation of China (Grant No.12365008) and the Guizhou Provincial Basic Research Program (Natural Science)(Grant No.QianKeHeJiChu-ZK[2024]YiBan027 and QianKeHeJiChuMS[2025]680).
 
%\nocite{*}
%\bibliographystyle{unsrt}
%\bibliography{ref}

\bibliography{ref}
\bibliographystyle{apsrev4-1}

\end{document}